\documentclass[showkeys,eqsecnum,twocolumn,showpacs,preprintnumbers,amssymb,aps]{revtex4}
\usepackage{graphicx}
\usepackage{dcolumn}
\usepackage{amsthm,amsmath}
\usepackage{longtable}

\begin{document}
\preprint{\today}

\title{Sc III Spectral Properties of Astrophysical Interest}
\vspace{0.5cm}

\author{D. K. Nandy, Yashpal Singh and B. K. Sahoo \footnote{Email: bijaya@prl.res.in}}
\affiliation{Theoretical Physics Division, Physical Research Laboratory, Ahmedabad-380009, India}
\author{Chengbin Li}
\affiliation{State Key Laboratory of Magnetic Resonance and Atomic and Molecular Physics, Wuhan Institute of Physics and Mathematics, Chinese Academy of Sciences,Wuhan 430071, China}

\vskip1.0cm

\begin{abstract}
Transition properties such as oscillator strengths, transition rates, 
branching ratios and lifetimes of many low-lying states in the doubly
ionized scandium (Sc III) are reported. A relativistic method in the
coupled-cluster framework has been employed to incorporate the electron
correlation effects due to the Coulomb interaction to all orders by considering
all possible singly and doubly excited electronic configurations conjointly
with the leading order triply excited configurations
in a perturbative approach.
Present results are compared with the previously reported results for the 
transition lines of astrophysical interest. In addition, some of the
transition properties and lifetimes of few low-lying states are given for the
first time. Role of the correlation effects in the evaluation of 
the transition strengths are described concisely.
\end{abstract} 


\maketitle

\section{Introduction}
The low-lying energy spectra of the doubly ionized scandium (Sc III) have been
studied precisely \cite{schippers,gibbs,smith,holmstroem}, however accurate 
results for other transition properties which are of astrophysical interest are
almost rarely investigated. Sc is one of the important elements available in the
photosphere of the sun \cite{lodders,zhang1,zhang2,francois}. With the accurate
information of the spectroscopic data of Sc and its ions, one can acquire
palpable knowledge about the abundance of this element in the solar
photosphere \cite{zhang1,zhang2}. Abundances of different elements in the sun
was studied latest by Anders et al \cite{anders}, but the Sc abundance is not 
well known yet in its photosphere. In that context, precise spectroscopic data
of Sc or its ions may be helpful for this purpose. These data
can also serve as reference to determine abundances of other elements
in the metal-poor stars \cite{zhang1}. From the variation study of the Sc
abundance pattern in the long lived F- and G- type stars with different
metallicity, it is possible to probe the nucleosynthesis and chemical 
evolution of the elements in our Galaxy \cite{zhang1,francois}. Ambiguity
in the finding of the overabundant of Sc in most of the metal rich stars
\cite{feltzing} can be resolved from its improved spectroscopic data. It is
also known that the collisional de-excitations of the metastable states
are rather slow which can lead to build-up of a population of metastable
levels due to M1 and E2 forbidden transitions both in the astrophysical objects
and primarily, in the low-density laboratory tokamak plasmas \cite{charro}. 
Intensities of these transitions are vital to infer knowledge about the
plasma temperature and dynamics which are of crucial quantities for the
determination of the electron density and temperature diagnostics in many
astronomical objects and in the laboratory tokamak plasmas \cite{charro}.

Sc III belongs to the potassium (K I ) isoelectronic sequence, but their energy
level schemes are different. Since Sc III is an ionized atomic system with
heavier nucleus than K I, it is expected that the orbitals of this ion are
more contracted towards the nucleus than the latter. Therefore, the electron 
correlation effects can be different in both the systems and the relativistic
effects in Sc III can be larger. Only a few rigorous calculations of transition
rates, oscillator strengths and lifetimes in a number of states in Sc III are
available till date and most of them are just using the mean-field theories. 
These theoretical, also observed in few cases, transition properties of Sc III 
can be found in \cite{buchta, andersen, zeippen, ali, sahoo}, out of which our
previous reported results on the transition rates and lifetimes of the 3d and 4s
states in this ion \cite{sahoo} were the latest. We had evaluated these 
quantities by calculating the forbidden transition amplitudes using the
relativistic coupled-cluster (RCC) method; an all order perturbative 
relativistic many-body approach. In the present work, we employ the same method
but account a large number of configuration interaction space to determine 
various transition properties of many low-lying states in the considered ion.
This method has also been employed successfully in other systems to study 
these properties very accurately \cite{bijaya1, bijaya2, bijaya3}. Some of
the Sc data through all the stages of ionization are being tabulated by
Wiese and Fuhr \cite{wiese} few decades ago and the corresponding Sc III data 
can be replaced by the results obtained from the present study for their uses 
in other applications; especially in the astrophysics.

 The remaining part of the paper is organized as follows: In the next 
section we describe the necessity of the oscillator strengths and lifetimes
for astrophysical studies along with the definitions of these quantities for
different multipole channels. Then we pursue with presenting and 
discussing the results in the following section before summarizing them.

\section{Theory and Method of calculations}
The emission coefficient from an upper level $k$ to the lower level $i$ 
in a given element for its diagnostic in an astronomical object is
given by \cite{goly}
\begin{eqnarray}
I_{ki} = \frac{2 \pi h e^2}{m_e } \frac{g_i f_{ik}}{ \lambda_{ki}} \frac{n}{u} \exp(-E_k/k_BT),
\label{eqn1}
\end{eqnarray}
where $\lambda_{ki}$, $f_{ik}$, $g_i$, $E_k$, $n$, $u$ and $T$ are the 
wavelength, absorption oscillator strength, statistical weight of the lower
level, energy of the upper level, particle density, partition function of an
atom or ion and excitation temperature, respectively. In the above expression
$h$, $e$, $m_e$ and $k_B$ are the universal constants. Therefore, accurate
values of $f_{ik}$ are necessary in order to identify the emission coefficients
$I_{ki}$ from different objects. It is also possible that $f_{ik}$ can be
extracted from the precisely observed $I_{ki}$ values and compared them with the
reported results to demonstrate the potency of the employed method. Moreover,
the temperature of an astrophysical object can be determined by plotting 
$ln \left ( \frac{I_{ki} \times \lambda_{ki}^3}{g_i f_{ik}} \right )$ against 
the $E_k$ values \cite{goly}.

  In the macroscopic mechanical equilibrium and with the knowledge of the gas 
density, the optical depth of the stellar atmosphere can be found by
\cite{dimitrijevic}
\begin{eqnarray}
\tau_{\lambda_{ki}} = \int_0^{\infty} d^3r {\cal V}_i \phi_{\lambda_{ki}} \frac{\pi e^2} {m_e c} f_{ik} \rho_i,
\label{eqn2}
\end{eqnarray}
where ${\cal V}_i$ is the volume density in the state $i$, $\phi_{\lambda_{ki}}$
 is the spectral line profile which can be obtained from the stellar 
atmosphere and $\rho_i$ is the gas density in the state $i$, respectively.
Accurate values of the oscillator strengths are also necessary for this purpose.

The emission (absorption) oscillator strength $f_{ki}$ ($f_{ik}$) is given
by \cite{sobelman}
\begin{eqnarray}
f_{ki} = 1.4992\times 10^{-16} A_{ki} \frac{g_k}{g_i}\lambda_{ki}^2
\label{eqn3}
\end{eqnarray}
where $\lambda_{ki}$ and the transition rate $A_{ki}$ are used
in \AA  and $s^{-1}$, respectively. Sometime the weighted oscillator strengths 
are commonly used which are obtained from the relation
\begin{eqnarray}
 g_i f_{ik} = - g_k f_{ki},
\label{eqn4}
\end{eqnarray}
with $g_i =(2 J_i+1)$, for $J$ being the angular momentum of the state.

The transition rates due to E1, E2 and M1 channels are given by
\begin{eqnarray}
A^{E1}_{ki} &=& \frac{64 \pi^4 e^2a_0^2}{3h\lambda_{ki}^3 g_k } = \frac{2.02613\times 10^{18}}{\lambda_{ki}^3 g_k} S_{ki}^{E1} \label{eqn5} \\
A^{E2}_{ki} &=& \frac{64 \pi^6 e^2a_0^4}{15h\lambda_{ki}^5 g_k} = \frac{1.11995\times 10^{18}}{\lambda_{ki}^5 g_k} S_{ki}^{E2} \label{eqn6} \\
\text{and} && \nonumber \\
A^{M1}_{ki} &=& \frac{64 \pi^4 e^2a_0^2 (\alpha/2)^2}{3h\lambda_{ki}^3 g_k} = \frac{2.69735\times 10^{13}}{\lambda_{ki}^3 g_k} S_{ki}^{M1}, \label{eqn7} \ \ \ \ \ \ \
\end{eqnarray}
respectively, where we are not accounting the transition rates due to
the $M2$ and $E3$ channels for their negligible magnitudes. In 
the above expressions, units of $A_{ki}$ and $\lambda_{ki}$ are maintained
with Eq. (\ref{eqn3}) and the line strengths are given in atomic unit 
(a.u.) for the corresponding channel $O$ which are defined as 
$S^O_{ki} = \mid {\langle J_k \vert \vert O \vert \vert J_i \rangle} \mid^2$.

The lifetime of a given state is estimated by taking reciprocal of the
total transition rates due to all possible channels $O$; i.e. the lifetime
of the state $k$ is given by
\begin{eqnarray}
\tau_k &=& \frac {1} {\sum_{O,i} A^{O}_{ki}}.
\label{eqn8}
\end{eqnarray}
Similarly, the branching ratio of a given transition in the channel $O$ from 
a state $k$ to a lower state $i$ is given by
\begin{eqnarray}
\Gamma^{O}_{ki} &=& \frac {A^{O}_{ki}} {\sum_{O,i} A^{O}_{ki}} \nonumber \\ 
 &=& \tau_k A^{O}_{k i}.
\label{eqn9}
\end{eqnarray}

  The considered ion Sc III has the ground state configuration as 
$[3p^6] \ 3d_{3/2}$ which can be separated into a closed-shell configuration 
$[3p^6]$ with the valence electron $3d_{3/2}$. By replacing $3d_{3/2}$ valence 
orbital with any excited state orbital in the above configuration, the
corresponding single excited states of this ion
can be obtained. In a Fock space representation, we assume a Fermi vacuum
as $\vert \Phi_0\rangle=[3p^6]$ and a reference state with a valence
orbital $v$ as $\vert \Phi_v \rangle= a_v^{\dagger} \vert \Phi_0\rangle$
to define different level of excitations. In this approach, it is customary
to express the atomic state function (ASF) in the (R)CC framework as
(e.g. see \cite{bijaya1,debasish})
\begin{eqnarray}
\vert \Psi_v \rangle &=& e^T \{1+S_v\} \vert \Phi_v \rangle , 
\label{eqn10}
\end{eqnarray}
where $T$ and $S_v$ represent the excitation operators carrying 
the core-core and core-valence electron correlation effects, respectively.
In this work, we consider all possible single and double excitations to
determine the amplitudes for the $T$ and $S_v$ operators and also the important
triple excitations are considered perturbatively in a self-consistent 
procedure only for the determination of the $S_v$ operator amplitudes; this
approach is known generally as the (R)CCSD(T) method.
Since Sc III is a medium size atomic system, the CCSD(T) method can be able
to incorporate the correlation effects in this ion comprehensively so that
the results can be obtained to the required precision.

Excitation amplitudes for $T$ operators are determined using the
equation
\begin{eqnarray}
\langle \Phi_0^* \vert \{\widehat{He^T}\} \vert \Phi_0 \rangle &=& 0,
\label{eqn11}
\end{eqnarray}
where $\vert \Phi_0^* \rangle$ represents all possible singly and doubly 
excited states with respect to $\vert \Phi_0 \rangle$. After obtaining these 
solutions, we obtain both the attachment energy $\Delta E_v$ (negative of 
the ionization 
potential (IP)) and $S_v$ amplitudes simultaneously for a given ASF of
configuration $[3p^6]$ with a valence electron denoted by $v$ using the 
equation
\begin{eqnarray}
\langle \Phi_v^L \vert \{\widehat{He^T}\} \{1+S_v\} \vert \Phi_v\rangle &=& \langle \Phi_v^L \vert 1+S_v \vert \Phi_v\rangle \times \nonumber \\ && \langle \Phi_v \vert \{\widehat{He^T}\} \{1+S_v\}  \vert \Phi_v\rangle \nonumber \\
 &=& \langle \Phi_v^L \vert \delta_{L,v}+S_v \vert \Phi_v\rangle \Delta E_v, \ \ \ \ \ \ \ \
\label{eqn12}
\end{eqnarray}
where the superscript $L$ represents for the singly ($L=1$) and doubly ($L=2$)
excited hole-particle states. The Dirac-Coulomb Hamiltonian has been 
considered for the present calculations.

The transition matrix element for a given channel $O$ from state $k$ to 
state $i$ is evaluated by calculating the expression
\begin{eqnarray}
\frac{\langle \Psi_k \vert O \vert \Psi_i \rangle}{\sqrt{\langle \Psi_k \vert \Psi_k\rangle \langle \Psi_i \vert \Psi_i\rangle}} &=& \frac{\langle \Phi_k \vert \{ 1+ S_k^{\dagger}\} \overline{ O} \{ 1+ S_i\} \vert \Phi_i\rangle}{\sqrt{ {\cal N}_k {\cal N}_i}}, \ \ \
\label{tab13}
\end{eqnarray}
where $\overline{O}=e^{T^{\dagger}}O e^T$ and ${\cal N}_v=\langle \Phi_v \vert 
\{ 1+ S_v^{\dagger} \} \overline{N} \{ 1+ S_v\} \vert \Phi_v\rangle$ with 
$\overline{N}=e^{T^{\dagger}} e^T$. These terms involve non-truncating series 
and their evaluation procedure is explained elsewhere, {e.g. see \cite{bijaya1,debasish}.

The trial DF wave function $\vert \Phi_0 \rangle$ is constructed initially 
using 32 Gaussian type orbitals (GTOs) for each angular momentum symmetry
before obtaining the self-consistent
solutions. To obtain the RCC wave functions, we have considered interaction 
space within 15s, 15p, 15d, 13f and 12g orbitals in contrast to 13s, 12p, 12d, 
7f and 5g orbitals in our previous work \cite{sahoo}.

\section{Results and Discussions}
We present first the IP results of various states from this work using the DF 
and CCSD(T) methods and compare them in Table \ref{tab1} with the corresponding 
values given in the NIST database \cite{nist}. The differences between the
CCSD(T) results and the NIST data are given as $\Delta$ in percentage in the
same table. As seen in the table, the differences between these results are 
sub-one per cent for all the states; in fact, most of the calculated results
are within half per cent accurate. Amount of the correlation effects in these
results annexed through the CCSD(T) method can be ascertained from the
differences between the DF and CCSD(T) results. Agreement between the
experimental results quoted in NIST database and CCSD(T) results signify
capability of the method for obtaining the correct results in the considered
system.

\begin{table}[h]
\caption{Ionization potentials of different states. Differences between the CCSD(T) and NIST results are given as $\Delta$.}
\begin{center}
\begin{tabular}{lcccc}
\hline \hline \\
State    &  DF & CCSD(T) & NIST \cite{nist} & $\Delta$ \\
         & (cm$^{-1}$)  & (cm$^{-1}$) &  (cm$^{-1}$) & (\%) \\
\hline
       &  & & \\
$3d \ ^2D_{3/2}$ & 186268.97 & 199168.89 & 199677.64 & 0.25 \\
$3d \ ^2D_{5/2}$ & 186104.28 & 198916.43 & 199479.73 & 0.28 \\
$4s \ ^2S_{1/2}$ & 168567.35 & 174283.19 & 174138.05 & $0.08$ \\
$4p \ ^2P_{1/2}$ & 133649.63 & 137631.36 & 137573.07 & 0.04 \\
$4p \ ^2P_{3/2}$ & 133205.60 & 136139.57 & 137099.19 & 0.70 \\
$4d \ ^2D_{3/2}$ &  50110.93 & 87392.72 & 87419.75  & 0.03 \\
$4d \ ^2D_{5/2}$ &  50089.33 & 87290.26 & 87374.42  & 0.10 \\
$5s \ ^2S_{1/2}$ &  83029.85 & 84743.04 & 84814.89  & 0.08 \\
$5p \ ^2P_{1/2}$ &  70102.15 & 71481.95 & 71570.25  & 0.12 \\
$5p \ ^2P_{3/2}$ &  69932.66 & 71299.70 & 71394.22 & 0.13 \\
$4f \ ^2F_{5/2}$ &  61959.64 & 62707.34 & 62803.50 & 0.15 \\
$4f \ ^2F_{7/2}$ &  61960.23 & 62707.42 & 62803.25 & 0.15 \\
$5d \ ^2D_{3/2}$ &  33000.09 & 51366.41 & 51547.34 & 0.35  \\
$5d \ ^2D_{5/2}$ &  32986.18 & 51342.51 & 51527.23 & 0.36 \\
$6s \ ^2S_{1/2}$ &  49524.46 & 50238.20 & 50483.34 & 0.79 \\         
$6p \ ^2P_{1/2}$ &  43206.41 & 43837.69 & 44187.59 & 0.79  \\
$6p \ ^2P_{3/2}$ &  43126.91 & 43752.33 & 44102.17 & 0.79 \\
\hline \hline \\
\end{tabular}
\end{center}
\label{tab1}
\end{table}
Although the calculated IP results seem to be accurate enough for considering
them in the {\it ab initio} determination of the transition properties, but it 
can be noticed that the errors associated in the energies get augmented in 
the estimation of the excitation energies (EEs); particularly between the fine 
structure states. This is because of the expected non-negligible contribution
from other higher relativistic corrections from the QED and Breit interactions
which are not considered in the present work. In contrast to the energies, the 
QED and Breit interaction contributions are known to be small in the 
estimation of the transition amplitudes. To minimize the uncertainties,
we use the experimental energies/wavelengths in the determination of other
transition properties.

In Table \ref{tab2}, we give the transition matrix elements including
their transition strengths due to the E1, M1 and E2 channels; other higher
order multiple channel contributions are very small to be neglected here.
These results can also be used to estimate the polarizabilities of different
states of the considered ion. As seen from the above table, among the forbidden
transitions the E2 transition amplitudes are generally significant except 
between the fine structure transitions where the M1 transition amplitudes are
also large enough to be accounted for. Role of the correlation effects  to determine 
these properties can be realized from the differences between the DF and 
CCSD(T) results given in the same table. Typically the magnitudes of the
amplitudes obtained using the CCSD(T) method are smaller compared to the
the DF results except where the results are minuscule. This cognition 
would be pertinent while we compare our transition rates, oscillator
strengths, branching ratios and lifetimes against the earlier reported
results which are obtained using the mean-field theory calculations.

\begin{center}
\begin{longtable}{lllll}
\caption{Calculated transition amplitudes and line strengths are given in a.u. for different channels.} \label{tab2}\\
\hline \hline
\multicolumn{2}{c}{Transition $i \rightarrow f$} & Dirac-Fock & CCSD(T) & $S_{i \rightarrow f}$ \\ \hline \\
\endfirsthead

\multicolumn{4}{r}{\tablename\ \thetable{} -- continuation from the previous table.} \\
\hline \hline
\multicolumn{2}{c}{Transition $i \rightarrow f$} & Dirac-Fock & CCSD(T) & $S_{i \rightarrow f}$ \\ \hline \\
\endhead

\hline
\multicolumn{4}{r}{ {\it Continue} \dots } \\
\endfoot

\hline \hline
\endlastfoot
$3d \ ^2D_{5/2}$ & $\xrightarrow{M1}3d \ ^2D_{3/2}$ & 1.549  & 1.541 & 2.37\\
                 & $\xrightarrow{E2}3d \ ^2D_{3/2}$ & 1.934 & 1.649 & 2.72\\

$4s \ ^2S_{1/2}$ & $\xrightarrow{M1}3d \ ^2D_{3/2}$ & $\sim 0$ & $-0.001$ & $\sim 0$ \\
                 & $\xrightarrow{E2}3d \ ^2D_{3/2}$ & 4.051 & 3.589 & 12.88\\
                 & $\xrightarrow{E2}3d \ ^2D_{5/2}$ & 4.975 & 4.414 & 19.48     \\

$4p \ ^2P_{1/2}$ & $\xrightarrow{E1}3d \ ^2D_{3/2}$  & $1.535$  & $1.325$ & 1.76  \\
                 & $\xrightarrow{E1}4s \ ^2S_{1/2}$  & $2.584$  & $2.345$ & 5.50   \\

$4p \ ^2P_{3/2}$ & $\xrightarrow{E1}3d \ ^2D_{3/2}$  & $0.683$  & $0.589$ & 0.35  \\
                 & $\xrightarrow{E1}3d \ ^2D_{5/2}$  & $-2.054$ & $-1.780$  & 3.17  \\
                 & $\xrightarrow{E1}4s \ ^2S_{1/2}$  & $-3.650$ & $-3.318$  & 11.01    \\
                 & $\xrightarrow{M1}4p \ ^2P_{1/2}$ &  $-1.154$  & $-1.154$ & 1.33    \\
                 & $\xrightarrow{E2}4p \ ^2P_{1/2}$ &$-12.452$ &$-11.713$   &137.19     \\
                                                                            
$4d \ ^2D_{3/2}$ & $\xrightarrow{M1}3d \ ^2D_{3/2}$  &  0.0002  & 0.0003    &$\sim 0$   \\
                 & $\xrightarrow{E2}3d \ ^2D_{3/2}$ &$-2.811$  &$-2.544$    &6.47   \\
                 & $\xrightarrow{M1}3d \ ^2D_{5/2}$ &  $-0.002$  & $-0.006$ &$\sim 0$     \\
                 & $\xrightarrow{E2}3d \ ^2D_{5/2}$ &$- 1.848$  &$- 1.678$  & 2.82     \\
                 & $\xrightarrow{M1}4s \ ^2S_{1/2}$ & $\sim 0$ & $\sim 0$   & $\sim 0$ \\
                 & $\xrightarrow{E2}4s \ ^2S_{1/2}$ &$-10.102$  &$-9.707$   &94.22     \\
                 & $\xrightarrow{E1}4p \ ^2P_{1/2}$  & $-3.907$ & $-3.719$  &13.83   \\  
                 & $\xrightarrow{E1}4p \ ^2P_{3/2}$  & $1.758$  &  $1.673$  & 2.80      \\
                                                                            
$4d \ ^2D_{5/2}$ & $\xrightarrow{M1}3d \ ^2D_{3/2}$ &  0.001   & 0.002      & $\sim 0$    \\
                 & $\xrightarrow{E2}3d \ ^2D_{3/2}$ & 1.837  & 1.662        & 2.76     \\
                 & $\xrightarrow{M1}3d \ ^2D_{5/2}$ &  0.0005  & 0.009      & $\sim 0$    \\
                 & $\xrightarrow{E2}3d \ ^2D_{5/2}$ & $-3.689$ & $-3.350$   & 11.22               \\
                 & $\xrightarrow{E2}4s \ ^2S_{1/2}$ & $-12.365$ & $-11.882$ &141.18                  \\
                 & $\xrightarrow{E1}4p \ ^2P_{3/2}$  & $5.270$  & $5.018$   &25.18                    \\  
                 & $\xrightarrow{M1}4d \ ^2D_{3/2}$ &  1.549   & 1.548      &2.40                 \\
                 & $\xrightarrow{E2}4d \ ^2D_{3/2}$ & 16.140 & 14.972       &224.16                      \\
                                                                            
$5s \ ^2S_{1/2}$ & $\xrightarrow{M1}3d \ ^2D_{3/2}$ & $\sim 0$ & $\sim 0$   &       $\sim 0$   \\
                 & $\xrightarrow{E2}3d \ ^2D_{3/2}$ & $-0.683$ & $-0.514$   &0.26                    \\
                 & $\xrightarrow{E2}3d \ ^2D_{5/2}$ & 0.844 & 0.643         &0.41                     \\
                 & $\xrightarrow{M1}4s \ ^2S_{1/2}$ &  $\sim 0$&$-0.002$    &$\sim 0$                      \\
                 & $\xrightarrow{E1}4p \ ^2P_{1/2}$ & $-1.453$ & $-1.442$   &2.08                       \\  
                 & $\xrightarrow{E1}4p \ ^2P_{3/2}$  & $-2.083$  & $-2.068$ &4.28                          \\
                 & $\xrightarrow{M1}4d \ ^2D_{3/2}$ & $\sim 0$ & $\sim 0$   &     $\sim 0$    \\
                 & $\xrightarrow{E2}4d \ ^2D_{3/2}$ & $-26.953$ & $-25.156$ &632.82                      \\
                 & $\xrightarrow{E2}4d \ ^2D_{5/2}$ & 33.052 & 30.872       &953.08                      \\
                                                                            
$5p \ ^2P_{1/2}$ & $\xrightarrow{E1}3d \ ^2D_{3/2}$ & $0.291$ & $0.251$     &0.06                    \\
                 & $\xrightarrow{E1}4s \ ^2S_{1/2}$ & $-0.106$ & $-0.179$   &0.03                        \\   
                 & $\xrightarrow{M1}4p \ ^2P_{1/2}$ &  $\sim 0$   & $\sim 0$         &   $\sim 0$          \\
                 & $\xrightarrow{M1}4p \ ^2P_{3/2}$ &$-0.005$   &$-0.005$   &$\sim 0$          \\  
                 & $\xrightarrow{E2}4p \ ^2P_{3/2}$ & $-7.734$ & $-7.403$   &54.80               \\
                 & $\xrightarrow{E1}4d \ ^2D_{3/2}$ & $4.578$ &  $4.330$    &18.75              \\    
                 & $\xrightarrow{E1}5s \ ^2S_{1/2}$ & $4.949$ & $4.849$     &23.51             \\    
                                                                            
$5p \ ^2P_{3/2}$ & $\xrightarrow{E1}3d \ ^2D_{3/2}$ & $-0.130$ & $-0.113$   &0.02               \\  
                 & $\xrightarrow{E1}3d \ ^2D_{5/2}$  & $0.392$ & $0.340$    &0.12                   \\    
                 & $\xrightarrow{E1}4s \ ^2S_{1/2}$  & $-0.132$ & $0.236$   &0.56                  \\    
                 & $\xrightarrow{M1}4p \ ^2P_{1/2}$ &  0.005   & 0.005      &$\sim 0$                   \\
                 & $\xrightarrow{E2}4p \ ^2P_{1/2}$ & $-7.540$ & $-7.209$   &51.97                  \\
                 & $\xrightarrow{M1}4p \ ^2P_{3/2}$ & $\sim 0$ & $\sim 0$   &       $\sim 0$      \\
                 & $\xrightarrow{E2}4p \ ^2P_{3/2}$ & $-7.660 $ & $-7.332$  &53.76                  \\
                 & $\xrightarrow{E1}4d \ ^2D_{5/2}$ & $-6.124$ & $-5.793$   &33.56                   \\    
                 & $\xrightarrow{E1}4d \ ^2D_{3/2}$  & $2.037$ & $1.936$    &3.75                  \\    
                 & $\xrightarrow{E1}5s \ ^2S_{1/2}$ & $7.063$  & $6.851$    &46.94                  \\    
                 & $\xrightarrow{M1}5p \ ^2P_{1/2}$ &  1.154   & 1.154      &1.33                  \\
                 & $\xrightarrow{E2}5p \ ^2P_{1/2}$ & 47.408 & 45.585       &2077.99               \\
                                                                            
$4f \ ^2F_{5/2}$ & $\xrightarrow{E1}3d \ ^2D_{3/2}$  & $-1.402$ & $-1.173$  &1.38                \\ 
                 & $\xrightarrow{E1}3d \ ^2D_{5/2}$  & $-0.376$ & $-0.315$  &0.011                   \\  
                 & $\xrightarrow{E2}4p \ ^2P_{1/2}$ & 17.580 & 16.611       &275.92                      \\
                 & $\xrightarrow{M1}4p \ ^2P_{3/2}$ & $\sim 0$ & $\sim 0$   &           $\sim 0$     \\
                 & $\xrightarrow{E2}4p \ ^2P_{3/2}$ & $-9.471$ & $-8.953$   &80.16                       \\
                 & $\xrightarrow{E1}4d \ ^2D_{3/2}$  & $-7.965$ & $-7.570$  &57.30                 \\
                 & $\xrightarrow{E1}4d \ ^2D_{5/2}$  & $-2.130$ & $-2.025$  &4.10                       \\
                 & $\xrightarrow{E2}5p \ ^2P_{1/2}$ & $-45.466$ & $-43.894$ &1926.68                  \\
                 & $\xrightarrow{M1}5p \ ^2P_{3/2}$ & $\sim 0$ & $\sim 0$   &            $\sim 0$     \\
                 & $\xrightarrow{E2}5p \ ^2P_{3/2}$ & $-24.318$ & $-23.480$ &551.31                        \\
                                                                            
$4f \ ^2F_{7/2}$ & $\xrightarrow{E1}3d \ ^2D_{5/2}$ & $-1.682$ & $-1.411$   &1.99                           \\
                 & $\xrightarrow{E2}4p \ ^2P_{3/2}$ & 23.12 & 23.20         &538.24                      \\
                 & $\xrightarrow{E1}4d \ ^2D_{5/2}$ & $-9.526$ & $-9.055$   &81.99                             \\
                 & $\xrightarrow{E2}5p \ ^2P_{3/2}$ & 59.16 & 59.56         &3547.40                            \\
                 & $\xrightarrow{M1}4f \ ^2F_{5/2}$ &  1.852 & 1.852        &3.43                          \\
                 & $\xrightarrow{E2}4f \ ^2F_{5/2}$ & 18.25 & 18.250        &333.06                           \\

$5d \ ^2D_{3/2}$ & $\xrightarrow{M1}3d \ ^2D_{3/2}$ &  0.0001  & 0.000      &0.0                          \\
                 & $\xrightarrow{E2}3d \ ^2D_{3/2}$ &$-0.976$ &$-0.917$     &0.84                       \\
                 & $\xrightarrow{M1}3d \ ^2D_{5/2}$ &  0.0008  & 0.003      &$\sim 0$                       \\
                 & $\xrightarrow{E2}3d \ ^2D_{5/2}$ &$-0.640$ &$-0.602$     &0.36                       \\
                 & $\xrightarrow{M1}4s \ ^2S_{1/2}$ & $\sim 0$ & $\sim 0$   &            $\sim 0$             \\
                 & $\xrightarrow{E2}4s \ ^2S_{1/2}$ &$-1.856$ &$-1.642$     &2.70                       \\
                 & $\xrightarrow{E1}4p \ ^2P_{1/2}$ & $-0.756$ & $-0.613$   &0.38                         \\
                 & $\xrightarrow{E1}4p \ ^2P_{3/2}$ & $0.334$  & $0.270$    &0.08                \\
                 & $\xrightarrow{M1}4d \ ^2D_{3/2}$ & $\sim 0$  & $-0.001$  &$\sim 0$                   \\
                 & $\xrightarrow{E2}4d \ ^2D_{3/2}$ &15.562  & 14.746       &217.44    \\    
                 & $\xrightarrow{M1}4d \ ^2D_{5/2}$ & 0.001 & 0.003         &$\sim 0$        \\      
                 & $\xrightarrow{E2}4d \ ^2D_{5/2}$ & 10.217 & 9.690        &93.90          \\     
                 & $\xrightarrow{M1}5s \ ^2S_{1/2}$ & $\sim 0$ & $\sim 0$   &        $\sim 0$   \\
                 & $\xrightarrow{E2}5s \ ^2S_{1/2}$ &$-31.762$ &$-31.420$   &987.22           \\
                 & $\xrightarrow{E1}5p \ ^2P_{1/2}$ & $-6.773$ & $-6.731$   &45.31          \\
                 & $\xrightarrow{E1}5p \ ^2P_{3/2}$ & $-3.048$ & $-3.030$   &9.19              \\
                 & $\xrightarrow{E1}4f \ ^2F_{5/2}$ & $-5.348$ & $-5.500$   &30.25               \\
                                                                            
$5d \ ^2D_{5/2}$ & $\xrightarrow{M1}3d \ ^2D_{3/2}$ & $-0.0007$  & 0.002    &$\sim 0$              \\
                 & $\xrightarrow{E2}3d \ ^2D_{3/2}$ & 0.638  & 0.560        &0.31               \\
                 & $\xrightarrow{M1}3d \ ^2D_{5/2}$ &  0.0003  & 0.0060     &$\sim 0$                 \\
                 & $\xrightarrow{E2}3d \ ^2D_{5/2}$ &$-1.280$&$-1.203$      &1.45                \\
                 & $\xrightarrow{E2}4s \ ^2S_{1/2}$ &$-2.280$ &$-2.010$     &4.04                  \\
                 & $\xrightarrow{E1}4p \ ^2P_{3/2}$ & $1.006$  & $0.813$    &0.66                     \\
                 & $\xrightarrow{M1}4d \ ^2D_{3/2}$ &  0.001   & 0.0003     &$\sim 0$                \\
                 & $\xrightarrow{E2}4d \ ^2D_{3/2}$ &$-10.168$ &$-9.635$    &92.83                        \\
                 & $\xrightarrow{M1}4d \ ^2D_{5/2}$ & $- 0.0002$  &$-0.0060$&$\sim 0$                   \\
                 & $\xrightarrow{E2}4d \ ^2D_{5/2}$ & 20.396 &19.344        &374.19                      \\
                 & $\xrightarrow{E2}5s \ ^2S_{1/2}$ &$-38.858$ &$-38.441$   &1477.71                     \\
                 & $\xrightarrow{E1}5p \ ^2P_{3/2}$ & $-9.138$ & $-9.082$   &82.48                     \\
                 & $\xrightarrow{E1}4f \ ^2F_{5/2}$ & $1.427$  & $1.468$    &2.16                          \\
                 & $\xrightarrow{E1}4f \ ^2F_{7/2}$ & $-6.382$ & $-6.564$   &43.09                        \\
                 & $\xrightarrow{M1}5d \ ^2D_{3/2}$ &  1.549   & 1.549      &2.40                         \\
                 & $\xrightarrow{E2}5d \ ^2D_{3/2}$ & 53.712 & 50.516       &2551.87                         \\
                                                                            
$6s \ ^2S_{1/2}$ & $\xrightarrow{M1}3d \ ^2D_{3/2}$ & $\sim 0$ &  $\sim 0$  &           $\sim 0$              \\
                 & $\xrightarrow{E2}3d \ ^2D_{3/2}$ & 0.237 &  0.193        &0.04              \\
                 & $\xrightarrow{E2}3d \ ^2D_{5/2}$ &$-0.292$ &$-0.240$     &0.06                \\
                 & $\xrightarrow{M1}4s \ ^2S_{1/2}$  & $\sim 0$ &  $0.002$  &$\sim 0$                   \\
                 & $\xrightarrow{E1}4p \ ^2P_{1/2}$  & $0.420$  &  $0.428$  &0.18                   \\
                 & $\xrightarrow{E1}4p \ ^2P_{3/2}$  & $0.598$  &  $0.614$  &0.38                     \\
                 & $\xrightarrow{M1}4d \ ^2D_{3/2}$ &$\sim 0$ &  $\sim 0$   &             $\sim 0$     \\
                 & $\xrightarrow{E2}4d \ ^2D_{3/2}$ &$-7.347$ & $-6.385$    &40.77                   \\ 
                 & $\xrightarrow{E2}4d \ ^2D_{5/2}$ & 9.039 &  7.871        &61.95               \\
                 & $\xrightarrow{M1}5s \ ^2S_{1/2}$  & $\sim 0$ &  $0.001$  &$\sim 0$                \\
                 & $\xrightarrow{E1}5p \ ^2P_{1/2}$  & $2.922$  &  $2.862$  &8.19               \\
                 & $\xrightarrow{E1}5p \ ^2P_{3/2}$  & $-4.182$ &  $-4.100$ &16.81                     \\
                 & $\xrightarrow{M1}5d \ ^2D_{3/2}$ &$\sim 0$  &$\sim 0$    &            $\sim 0$         \\
                 & $\xrightarrow{E2}5d \ ^2D_{3/2}$ &86.375  & 81.81        &6692.88                  \\
                 & $\xrightarrow{E2}5d \ ^2D_{5/2}$ &$-105.891$ &$-100.318$ &10063.70               \\
                                                                            
$6p \ ^2P_{1/2}$ & $\xrightarrow{E1}3d \ ^2D_{3/2}$  & $-0.152$ &  $-0.128$ &0.02                  \\
                 & $\xrightarrow{E1}4s \ ^2S_{1/2}$  & $0.068$  &  $0.115$  &0.01                   \\
                 & $\xrightarrow{M1}4p \ ^2P_{1/2}$ &   $\sim 0$  & $\sim 0$&      $\sim 0$             \\
                 & $\xrightarrow{M1}4p \ ^2P_{3/2}$ &   0.002  & 0.003      &$\sim 0$                  \\
                 & $\xrightarrow{E2}4p \ ^2P_{3/2}$ &   2.118 &   2.074     &4.30                  \\
                 & $\xrightarrow{E1}4d \ ^2D_{3/2}$  & $0.498$  &  $0.512$  &0.26                    \\
                 & $\xrightarrow{E1}5s \ ^2S_{1/2}$  & $0.054$  &  $0.093$  &0.01                  \\
                 & $\xrightarrow{M1}5p \ ^2P_{1/2}$ &   $\sim 0$  & $\sim 0$&         $\sim 0$           \\
                 & $\xrightarrow{M1}5p \ ^2P_{3/2}$ & $-0.005$  &$-0.005$   &$\sim 0$                    \\
                 & $\xrightarrow{E2}5p \ ^2P_{3/2}$ & $-24.547$ &$-23.634$  &558.57                    \\
                 & $\xrightarrow{E2}4f \ ^2F_{5/2}$ & $-18.899$ & $-18.629$ &347.04                     \\ 
                 & $\xrightarrow{E1}5d \ ^2D_{3/2}$  & $8.600$  &  $8.168$  &66.72                   \\
                 & $\xrightarrow{E1}6s \ ^2S_{1/2}$  & $8.268$  &  $8.160$  &66.58                     \\

$6p \ ^2P_{3/2}$ & $\xrightarrow{E1}3d \ ^2D_{3/2}$  & $-0.068$ &  $-0.057$ &0.003                     \\
                 & $\xrightarrow{E1}3d \ ^2D_{5/2}$  & $0.205$  &  $0.174$  &0.03                  \\
                 & $\xrightarrow{E1}4s \ ^2S_{1/2}$  & $-0.088$ &  $-0.155$ &0.02                    \\
                 & $\xrightarrow{M1}4p \ ^2P_{1/2}$ &   0.002  & 0.002      &$\sim 0$                 \\
                 & $\xrightarrow{E2}4p \ ^2P_{1/2}$ &$-2.092$ &$-2.048$     &4.19                   \\
                 & $\xrightarrow{M1}4p \ ^2P_{3/2}$ &$\sim 0$    &$\sim 0$  &             $\sim 0$         \\
                 & $\xrightarrow{E2}4p \ ^2P_{3/2}$ &  2.110  &   2.068     &4.28                    \\
                 & $\xrightarrow{E1}4d \ ^2D_{3/2}$  & $0.226$  &  $ 0.232$ &0.05                   \\
                 & $\xrightarrow{E1}4d \ ^2D_{5/2}$  & $-0.678$ &  $-0.696$ &0.48                    \\
                 & $\xrightarrow{E1}5s \ ^2S_{1/2}$  & $-0.048$ &  $-0.102$ &0.01                     \\
                 & $\xrightarrow{M1}5p \ ^2P_{1/2}$ &   0.005  & 0.005      &$\sim 0$                  \\
                 & $\xrightarrow{E2}5p \ ^2P_{1/2}$ &  $-23.928$ &$-22.998$ &528.90                       \\
                 & $\xrightarrow{M1}5p \ ^2P_{3/2}$ & $\sim 0$  &$0.003$    &$\sim 0$                  \\
                 & $\xrightarrow{E2}5p \ ^2P_{3/2}$ & $24.295$ &$23.368$    &546.06                     \\
                 & $\xrightarrow{M1}4f \ ^2F_{5/2}$ & $\sim 0$ & $\sim 0$   &                $\sim 0$      \\
                 & $\xrightarrow{E2}4f \ ^2F_{5/2}$ & $-9.998$ & $-9.852$   &97.06                     \\
                 & $\xrightarrow{E2}4f \ ^2F_{7/2}$ & $24.489$ & $24.131$   &582.26                         \\
                 & $\xrightarrow{E1}5d \ ^2D_{3/2}$  & $3.830$  &  $ 3.636$ &13.22                      \\
                 & $\xrightarrow{E1}5d \ ^2D_{5/2}$  & $-11.506$&  $-10.928$&119.42                     \\
                 & $\xrightarrow{E1}6s \ ^2S_{1/2}$  & $-11.676$&  $-11.522$&132.76                      \\
                 & $\xrightarrow{M1}6p \ ^2P_{1/2}$ &  $- 1.154$  &$- 1.154$&1.33                  \\
                 & $\xrightarrow{E2}6p \ ^2P_{1/2}$ &$-133.050$  & $-129.491$&16641.00                 \\
\end{longtable}                                                              
\end{center}                                                               
Using the transition amplitudes/strengths from the CCSD(T) calculations given in Table \ref{tab2} 
and experimental wavelengths estimated from the NIST database energies (given 
in Table \ref{tab1}), we determine the transition rates, emission oscillator
strengths and branching ratios of various transitions and present them in Table
\ref{tab3}. We have also compared our results with other available results
for these properties. There are also few calculations available on the
transition probabilities and oscillator strengths earlier. Transition 
probabilities reported by us in our earlier work \cite{sahoo} which were
obtained using the same method of the present work, however considering a
larger size of configuration interaction space with the availability of a bigger
computational resource in order to verify the consistent of the results with
respect to our previous calculations. We find the results are still consistent
with our previous calculations. Ali and Kim have also calculated these
forbidden transition rates \cite{ali} using the multi-configuration Dirac-Fock 
(MCDF) method, their results are also in agreement with us except for the
M1 amplitude of the $4s \ ^2S_{1/2} \rightarrow 3d \ ^2D_{3/2}$ transition.
In fact the MCDF method is incompetent to account the correlation effects as
comprehensively as the RCC method, especially the core-polarization
correlations. From our calculations we observe that the above M1 amplitude
is about $5.12 \times 10^{-6}$ at the DF level and the core-polarization
effects through the core correlations aggrandize it to be $-0.001$ (a.u.) in 
the CCSD(T) method. This is the main reason for the discrepancy
between the results obtained from these two methods and it advocates for the
essence of studying the transition properties using a method like
our RCC approach. In another work, Zeippen has also employed the SUPERSTRUCTURE
program to estimate these forbidden transition rates besides for some
other ions by scaling the wave functions and energies. In that work the
results are also compared with the above results of Ali and Kim except for
the above discussed M1 transition amplitude which is not reported at all. Our 
results also agree reasonably well with their calculations. In 1975, Wiese 
and Fuhr have tabulated most of the transition rates and oscillator
strengths due to the allowed transitions accumulating from various works
\cite{wiese}. The calculated results reported in their list were obtained
from the non-relativistic mean-field methods and other results were 
taken from the observations. Most of our results are comparable with the 
tabulated results, however the present calculations are believed to be
meticulous than those are tabulated in the above reference. This may be
perceptible while one scrutinizes the discussions in the next paragraph. Along
with the results discussed above, we also present the forbidden transition 
properties for all these transitions although their contributions seem 
to be irrelevant for the determination of the lifetimes of the excited states 
those are considered except for the first two (it will be evident later). However,
these results could be useful for some other purposes like estimating the higher
multipole polarizabilities, Stark shifts etc. Also, the transition properties
for the $6s$ and $6p$ states were not known earlier. We also give the branching 
ratios of all the transitions in the same table when their values are
significant up to three decimal places. It is possible by us to estimate these
values due to determination all the important transition rates in this work.

\begin{longtable*}{lclllllc}
\caption{Wavelengths ($\lambda$ in \AA), transition rates ($A$ in $s^{-1}$), oscillator strengths ($f$) and branching ratios ($\Gamma$) from different works. Numbers given as $[k]$ implies $\times 10^k$. } \label{tab3}\\
\hline \hline
Upper & Lower & $\lambda_{f \rightarrow i}$ & \multicolumn{2}{c}{$A^{\text{O}}_{f \rightarrow i}$} &  \multicolumn{2}{c}{$f$} & $\Gamma$\\
state ($f$) &  state ($i$) &  & Others & Present & Others & Present & Present\\
\hline\\
\endfirsthead

\multicolumn{7}{c}{\tablename\ \thetable{} -- continued from previous page.} \\
\hline \hline
Upper & Lower & $\lambda_{f \rightarrow i}$ & \multicolumn{2}{c}{$A^{\text{O}}_{f \rightarrow i}$} &  \multicolumn{2}{c}{$f$} & $\Gamma$\\
state ($f$) &  state ($i$) &  & Others & Present & Others & Present & Present\\
\hline\\
\endhead

\hline
\multicolumn{7}{c}{ {\it Continue} \dots } \\
\endfoot

\hline \hline
\endlastfoot
$3d_{5/2}$  & $\xrightarrow{M1}3d_{3/2}$ & 505970.4 & 8.32[-5]$^a$  &  8.33[-5]   & & $\sim 0$  & $\sim 1.0$  \\ 
         &                              &          & 8.32[-5]$^b$    &   &  \\
         &                              &          & 8.24[-5]$^c$    &   &  \\
            & $\xrightarrow{E2}3d_{3/2}$ &        & 1.75[-11]$^b$ &  1.53[-11]   & & $\sim 0$ & $\sim 0.0$   \\
         &                              &          & 1.53[-11]$^c$    &   &  \\

$4s_{1/2}$ & $\xrightarrow{M1}3d_{3/2}$ & 3915.53   & 1.05[-8]$^b$  &  1.95[-4]    & & $\sim 0$ &  $\sim 0.0$ \\
         &                              &          & 1.79[-4]$^c$    &   &   \\
         & $\xrightarrow{E2}3d_{3/2}$   &          & 7.95$^a$    &  7.83   & & $\sim 0$ & 0.407 \\
         &                              &          & 8.21$^b$    &   &   \\
         &                              &          & 7.86$^c$    &   &   \\
         & $\xrightarrow{E2}3d_{5/2}$   & 3946.07   & 11.5$^a$    & 11.40        & & $\sim 0$ & 0.593  \\
         &                              &          & 11.9$^b$    &   &   \\
         &                              &          & 11.41$^c$    &   &  \\

$4p_{1/2}$ & $\xrightarrow{E1}3d_{3/2}$ & 1610.194  &  4.4[8]$^d$ &  4.26[8]     & 0.085$^d$ & 0.083 & 0.389 \\
         & $\xrightarrow{E1}4s_{1/2}$   & 2734.857  & 3.3[8]$^d$   & 2.72[8]      & 0.37$^d$ & 0.305   & 0.610  \\

$4p_{3/2}$ & $\xrightarrow{E1}3d_{3/2}$ & 1598.00   &  4.6[7]$^d$  & 4.31[7]     & 0.018$^d$ & 0.017  &  0.060 \\
         & $\xrightarrow{E1}3d_{5/2}$   & 1603.06   &  4.1[8]$^d$    & 3.90[8]      & 0.10$^d$ & 0.100  & 0.544   \\
         & $\xrightarrow{E1}4s_{1/2}$   & 2699.87   & 3.3[8]$^d$   & 2.83[8]      & 0.73$^d$ & 0.618  & 0.395   \\
         & $\xrightarrow{M1}4p_{1/2}$   & 211023.9  &   & 9.56[-4]     & & $\sim 0$ & $\sim 0.0$ \\
         & $\xrightarrow{E2}4p_{1/2}$   &         &   & 9.18[-8]     & & $\sim 0$ & $\sim 0.0$ \\

$4d_{3/2}$ & $\xrightarrow{M1}3d_{3/2}$ &  890.81  &   & 7.95[-4]    & & $\sim 0$ &  $\sim 0.0$\\
        & $\xrightarrow{E2}3d_{3/2}$    &          &    & 3.23[3]     & & $\sim 0$ & $\sim 0.0$  \\
        & $\xrightarrow{M1}3d_{5/2}$    &  892.38  &      & 0.366       & & $\sim 0$ & $\sim 0.0$ \\
        & $\xrightarrow{E2}3d_{5/2}$    &       &   & 1.39[3]    & & $\sim 0$ & $\sim 0.0$ \\
        & $\xrightarrow{M1}4s_{1/2}$    & 1153.16  &  & 2.80[-4]    & & $\sim 0$ & $\sim 0.0$  \\
        & $\xrightarrow{E2}4s_{1/2}$    &       &    & 1.29[4]     & & $\sim 0$ & $\sim 0.0$ \\
        & $\xrightarrow{E1}4p_{1/2}$    & 1993.89  &  9.6[8]$^d$  & 8.81[8]     &  1.1$^d$ & 1.050  & 0.825   \\
        & $\xrightarrow{E1}4p_{3/2}$    & 2012.91  & 1.9[8]$^d$  & 1.74[8]  &  0.11$^d$ & 0.106  & 0.175  \\

$4d_{5/2}$ & $\xrightarrow{M1}3d_{3/2}$ &  890.45   &    & 0.003        & & $\sim 0$  & $\sim 0.0$ \\
        & $\xrightarrow{E2}3d_{3/2}$    &         &   & 921.285      & & $\sim 0$ & $\sim 0.0$  \\
        & $\xrightarrow{M1}3d_{5/2}$    &  892.02   &     & 0.531        & & $\sim 0$  & $\sim 0.0$ \\
        & $\xrightarrow{E2}3d_{5/2}$    &        &    & 3.71[3]      & & $\sim 0$ & $\sim 0.0$ \\
        & $\xrightarrow{E2}4s_{1/2}$    &  1152.56  &    & 1.30[4]      & & $\sim 0$ & $\sim 0.0$ \\
        & $\xrightarrow{E1}4p_{3/2}$    &  2011.07   & 1.1[9]$^d$   & 1.05[9]  & 1.0$^d$  & 0.955  & $\sim 1.0$  \\
        & $\xrightarrow{M1}4d_{3/2}$    & 2206044.0 &   & 1.00[-6]     & & $\sim 0$ & $\sim 0.0$ \\
        & $\xrightarrow{E2}4d_{3/2}$    &      &  & 8.00[-13]    & & $\sim 0$ & $\sim 0.0$ \\

$5s_{1/2}$ & $\xrightarrow{M1}3d_{3/2}$    & 870.61   &  & 2.87[-3]     & & $\sim 0$ & $\sim 0.0$ \\
        & $\xrightarrow{E2}3d_{3/2}$       &    &  & 296.429      & & $\sim 0$ & $\sim 0.0$ \\
        & $\xrightarrow{E2}3d_{5/2}$       & 872.11   &   & 458.998      & & $\sim 0$ & $\sim 0.0$ \\
        & $\xrightarrow{M1}4s_{1/2}$       & 1119.53  &  &5.54[-2]      & & $\sim 0$  & $\sim 0.0$   \\
        & $\xrightarrow{E1}4p_{1/2}$       & 1895.44  &  2.8[8]$^d$ & 3.12[8] & 0.15$^d$ & 0.168    & 0.350   \\
        & $\xrightarrow{E1}4p_{3/2}$       & 1912.62  & 5.4[8]$^d$ & 5.88[8]      & 0.15$^d$ & 0.161  & 0.653   \\
        & $\xrightarrow{M1}4d_{3/2}$       & 38389.78 &  &3.43[-9]      & & $\sim 0$ & $\sim 0.0$   \\
        & $\xrightarrow{E2}4d_{3/2}$       &    &  &4.25[-3]  & & $\sim 0$ & $\sim 0.0$ \\
        & $\xrightarrow{E2}4d_{5/2}$       & 39069.67 &  & 5.86[-3]     & & $\sim 0$ & $\sim 0.0$ \\

$5p_{1/2}$ & $\xrightarrow{E1}3d_{3/2}$ & 780.60  & 1.5[8]$^d$  & 1.35[8]  & 0.0066$^d$ & 0.006  & 0.448  \\
        & $\xrightarrow{E1}4s_{1/2}$    &  974.97    &    & 3.51[7]      & & 0.005  & 0.116  \\
       & $\xrightarrow{M1}4p_{1/2}$     &  1515.09  &   & $\sim 0$   & & $\sim 0$ & $\sim 0$\\
        & $\xrightarrow{M1}4p_{3/2}$    & 1526.04  &     & 0.113       & & $\sim 0$ & $\sim 0$ \\
        & $\xrightarrow{E2}4p_{3/2}$    &        &   & 3.71[3]      & & $\sim 0$ & $\sim 0$ \\
        & $\xrightarrow{E1}4d_{3/2}$    &  6309.35   &  7.0[7]$^d$  & 7.56[7]      & 0.21$^d$ & 0.226 & 0.251 \\
        & $\xrightarrow{E1}5s_{1/2}$    & 7550.22    & 5.4[7]$^d$   & 5.53[7]      & 0.47$^d$ & 0.473 & 0.184 \\

$5p_{3/2}$ & $\xrightarrow{E1}3d_{3/2}$ &  779.53   & 1.5[7]$^d$  & 1.38[7]   & 0.0014$^d$ & 0.001 & 0.046 \\
       & $\xrightarrow{E1}3d_{5/2}$     &   780.73   &  1.3[8]$^d$  & 1.23[8] & 0.0079$^d$  & 0.007  & 0.407  \\ 
       & $\xrightarrow{E1}4s_{1/2}$     &  973.29   &    & 3.07[7]    & & 0.009   & 0.101  \\
       & $\xrightarrow{M1}4p_{1/2}$     &  1511.06  &     & 0.005      & & $\sim 0$ & $\sim 0.0$\\
       & $\xrightarrow{E2}4p_{1/2}$     &          &    & 1.85[3]    & & $\sim 0$  & $\sim 0.0$ \\
       & $\xrightarrow{M1}4p_{3/2}$     &  1521.95   &   & 5.92[-5]   && $\sim 0$& $\sim 0.0$   \\
       & $\xrightarrow{E2}4p_{3/2}$     &          &    & 1.84[3]    & & $\sim 0$& $\sim 0.0$   \\
       & $\xrightarrow{E1}4d_{3/2}$     &  6240.04  &  7.72[6]$^d$  & 7.74[6]    & 0.042$^d$ & 0.045 & 0.026   \\
       & $\xrightarrow{E1}4d_{5/2}$     &   6257.74  & 6.5[7]$^d$   & 6.94[7]    & 0.25$^d$ & 0.272 & 0.229 \\
       & $\xrightarrow{E1}5s_{1/2}$     &   7451.19  &  5.7[7]$^d$ & 5.75[7]    & 0.94$^d$& 0.957   & 0.190  \\
       & $\xrightarrow{M1}5p_{1/2}$     &  568085.0  &   & 4.90[-5]   & & $\sim 0$ & $\sim 0$ \\
       & $\xrightarrow{E2}5p_{1/2}$     &            &   & 9.83[-9]   & & $\sim 0$ & $\sim 0$ \\
       & $\xrightarrow{M1}5p_{1/2}$  &  1521.96   &   & 5.92[-5]   & & $\sim 0$ & $\sim 0$  \\
       & $\xrightarrow{E2}5p_{1/2}$  &            &    & 1.84[3]    & & $\sim 0$ & $\sim 0$ \\

$4f_{5/2}$ & $\xrightarrow{E1}3d_{3/2}$  & 730.60    &  1.1[9]$^d$  & 1.19[9]  & 0.13$^d$ & 0.143   & 0.751   \\
           & $\xrightarrow{E1}3d_{5/2}$  &  731.66   & 7.8[7]$^d$   & 8.59[7]    & 0.0062$^d$ & 0.007  & 0.051   \\
           & $\xrightarrow{E2}4p_{1/2}$  &  337.443   &    & 1.20[4]    & & $\sim 0$ & $\sim 0$ \\
           & $\xrightarrow{M1}4p_{3/2}$  &  1345.97   &   & 2.56[-6]   & & $\sim 0$ & $\sim 0$ \\
           & $\xrightarrow{E2}4p_{3/2}$  &        &    & 3.39[3]    & & $\sim 0$ & $\sim 0$  \\
           & $\xrightarrow{E1}4d_{3/2}$  &   4062.36   & 2.9[8]$^d$   & 2.89[8]    & 1.1$^d$ & 1.072 & 0.182     \\
           & $\xrightarrow{E1}4d_{5/2}$  &   4069.85   & 2.1[7]$^d$   & 2.05[7]    & 0.052$^d$ & 0.051  & 0.013   \\
           & $\xrightarrow{E2}5p_{1/2}$  & 11406.74  &      & 1.862      & & $\sim 0$ & $\sim 0$ \\
           & $\xrightarrow{M1}5p_{3/2}$  & 11640.47  &  & 1.35[-11]  & & $\sim 0$ & $\sim 0$ \\
           & $\xrightarrow{E2}5p_{3/2}$  &    &    & 0.481    &  & $\sim 0$ & $\sim 0$ \\

$4f_{7/2}$ & $\xrightarrow{E1}3d_{5/2}$  & 731.65     &  1.1[9]$^d$   & 1.29[9]     & 0.12$^d$ & 0.138  & 0.807   \\
           & $\xrightarrow{E2}4p_{3/2}$  & 1345.97    &     & 1.71[4]     & & $\sim 0$ & $\sim 0$ \\
           & $\xrightarrow{E1}4d_{5/2}$  & 4069.81    & 3.1[8]$^d$    & 3.08[8]     & 1.0$^d$ & 1.021 & 0.192  \\
           & $\xrightarrow{E2}5p_{3/2}$  & 11640.13   &       & 2.323       & & $\sim 0$ & $\sim 0$ \\
           & $\xrightarrow{M1}4f_{5/2}$  & 4.00[8]    &   & 1.81[-13]   & & $\sim 0$  & $\sim 0$ \\
           & $\xrightarrow{E2}4f_{5/2}$  &            &   & 4.55[-24]   & & $\sim 0$  & $\sim 0$ \\

$5d_{3/2}$ & $\xrightarrow{M1}3d_{3/2}$  & 676.58  &   & 3.46[-3]    & & $\sim 0$  & $\sim 0$ \\
           & $\xrightarrow{E2}3d_{3/2}$  &     &    & 1.66[3]     & & $\sim 0$  & $\sim 0$\\
           & $\xrightarrow{M1}3d_{5/2}$  & 1954.32 &      & 0.001       & & $\sim 0$ & $\sim 0$\\
           & $\xrightarrow{E2}3d_{5/2}$  & 677.74  &    &  710.019    & & $\sim 0$ & $\sim 0$ \\
           & $\xrightarrow{M1}4s_{1/2}$  & 827.02  &   &  2.93[-4]   &  & $\sim 0$ & $\sim 0$\\
           & $\xrightarrow{E2}4s_{1/2}$  &     &   &  1.95[3]    &  & $\sim 0$ & $\sim 0$\\
           & $\xrightarrow{E1}4p_{1/2}$  & 1159.22 & 1.6[8]$^d$   & 1.22[8]     & 0.067$^d$ & 0.050    & 0.312    \\
           & $\xrightarrow{E1}4p_{3/2}$  & 1179.62 & 3.2[7]$^d$   & 2.25[7]     & 0.0066$^d$ & 0.005  &  0.057  \\
           & $\xrightarrow{M1}4d_{3/2}$  & 2775.75 & $\sim 0$    &     &  $\sim 0$ & $\sim 0$ \\
           & $\xrightarrow{E2}4d_{3/2}$  &    &    &  369.313    & & $\sim 0$ & $\sim 0$ \\
           & $\xrightarrow{M1}4d_{5/2}$  & 2783.66 &   &  2.09[-4]   & & $\sim 0$ & $\sim 0$ \\
           & $\xrightarrow{E2}4d_{5/2}$  & 2783.67 &    &  157.227    & & $\sim 0$ & $\sim 0$ \\
           & $\xrightarrow{M1}5s_{1/2}$  & 2996.10 &   &  9.03[-7]   & & $\sim 0$ & $\sim 0$\\
           & $\xrightarrow{E2}5s_{1/2}$  &    &   & 1.14[3]    &  & $\sim 0$ & $\sim 0$\\
           & $\xrightarrow{E1}5p_{1/2}$  & 4971.28 & 1.8[8]$^d$   & 1.87[8]     & 1.4$^d$ & 1.386 & 0.479    \\
           & $\xrightarrow{E1}5p_{3/2}$  & 5016.73 &  3.6[7]$^d$  & 3.68[7]     & 0.14$^d$ & 0.139   & 0.094  \\
           & $\xrightarrow{E1}4f_{5/2}$  & 8817.62 & 2.1[7]$^d$   & 2.23[7]     & 0.16$^d$ & 0.173   & 0.057 \\

$5d_{5/2}$  & $\xrightarrow{M1}3d_{3/2}$  &  674.99   &     & 0.004     & & $\sim 0$  & $\sim 0$ \\
            & $\xrightarrow{E2}3d_{3/2}$  &       &  & 479.243   & & $\sim 0$ & $\sim 0$ \\
            & $\xrightarrow{M1}3d_{5/2}$  & 675.89    &    & 0.494     & & $\sim 0$ & $\sim 0$\\
            & $\xrightarrow{E2}3d_{5/2}$  &       &   & 2.2[3]    & & $\sim 0$ & $\sim 0$\\
            & $\xrightarrow{E2}4s_{1/2}$  & 815.59    &   & 2.11[3]   & & $\sim 0$ & $\sim 0$ \\
            & $\xrightarrow{E1}4p_{3/2}$  & 1168.61   & 1.9[8]$^d$  & 1.40[8]   & 0.060$^d$ & 0.043  & 0.368   \\
            & $\xrightarrow{M1}4d_{3/2}$  & 2786.09   &  & 1.50[-5]  & & $\sim 0$ & $\sim 0$\\
            & $\xrightarrow{E2}4d_{3/2}$  &     &   & 103.186   & & $\sim 0$ & $\sim 0$ \\
            & $\xrightarrow{M1}4d_{5/2}$  & 2789.62   &  & 6.55[-3]  & & $\sim 0$ & $\sim 0$ \\
            & $\xrightarrow{E2}4d_{5/2}$  &       &   & 413.268   & & $\sim 0$  & $\sim 0$\\
            & $\xrightarrow{E2}5s_{1/2}$  & 3004.12   &   & 1.13[3]   & & $\sim 0$ & $\sim 0$ \\
            & $\xrightarrow{E1}5p_{3/2}$  & 5033.47   &  2.2[8]$^d$ & 2.18[8]   & 1.2$^d$ & 1.242 & 0.573 \\
            & $\xrightarrow{E1}4f_{5/2}$  & 8868.18   & 0.99[6]$^d$  & 1.04[6]   & 0.012$^d$ & 0.012 & 0.023 \\
            & $\xrightarrow{E1}4f_{7/2}$  & 8868.38   & 2.0[7]$^d$  & 2.09[7]   & 0.18$^d$ & 0.185  & 0.055  \\
            & $\xrightarrow{M1}5d_{3/2}$  & 4972650.0 &  & 8.78[-8]  & & $\sim 0$ & $\sim 0$ \\
            & $\xrightarrow{E2}5d_{3/2}$  &       & & 1.57[-13] & & $\sim 0$ & $\sim 0$ \\

$6s_{1/2}$  & $\xrightarrow{M1}3d_{3/2}$  & 670.27   &  & 2.07[-3] & & $\sim 0$ & $\sim 0$ \\
            & $\xrightarrow{E2}3d_{3/2}$  &    &  & 1.54[2]  & & $\sim 0$ & $\sim 0$ \\
            & $\xrightarrow{E2}3d_{5/2}$  & 671.16    &  & 237.108  & & $\sim 0$ & $\sim 0$  \\
            & $\xrightarrow{M1}4s_{1/2}$  & 808.70    &  & 6.53[-2] & & $\sim 0$ &  $\sim 0$   \\
            & $\xrightarrow{E1}4p_{1/2}$  & 1148.24   &   & 1.23[8]  & & 0.024  & 0.201 \\
            & $\xrightarrow{E1}4p_{3/2}$  & 1154.52   &   & 2.48[8]  & & 0.025 & 0.406 \\
            & $\xrightarrow{M1}4d_{3/2}$  & 2707.36   &  & 6.80[-6] & & $\sim 0$     & $\sim 0$      \\
            & $\xrightarrow{E2}4d_{3/2}$  &   &   & 156.908  & & $\sim 0$ & $\sim 0$ \\
            & $\xrightarrow{E2}4d_{5/2}$  & 2710.68   &   & 236.973  & & $\sim 0$ & $\sim 0$ \\
            & $\xrightarrow{M1}5s_{1/2}$  & 2912.77   &  & 5.46[-4] & & $\sim 0$ & $\sim 0$  \\
            & $\xrightarrow{E1}5p_{1/2}$  & 4742.28   &   & 7.78[7]  & & 0.262  & 0.127  \\
            & $\xrightarrow{E1}5p_{3/2}$  & 4782.20   &   & 1.62[8]  & & 0.278  & 0.265 \\
            & $\xrightarrow{M1}5d_{3/2}$  & 93984.96  &  & 3.09[-11] & & $\sim 0$ & $\sim 0$\\
            & $\xrightarrow{E2}5d_{3/2}$  &     &  & 5.11[-4]  & & $\sim 0$ & $\sim 0$ \\
            & $\xrightarrow{E2}5d_{5/2}$  & 95795.53  &  & 6.98[-4]  & & $\sim 0$ & $\sim 0$ \\

$6p_{1/2}$ & $\xrightarrow{E1}3d_{3/2}$   &  643.13  &   & 6.27[7]  & & 0.002     & 0.397  \\
           & $\xrightarrow{E1}4s_{1/2}$   & 769.52  &   &  2.96[7] & & 0.003     & 0.187   \\
           & $\xrightarrow{M1}4p_{1/2}$   & 1070.83 &   & 4.40[-4] & & $\sim 0$ & $\sim 0$  \\
           & $\xrightarrow{M1}4p_{3/2}$   & 1076.29 &   &  0.009   & & $\sim 0$  & $\sim 0$  \\
           & $\xrightarrow{E2}4p_{3/2}$   &     &    &  1.67[3] & & $\sim 0$ & $\sim 0$ \\
           & $\xrightarrow{E1}4d_{3/2}$   & 2313.09 &    & 2.15[7]  & & 0.009    & 0.136  \\
           & $\xrightarrow{E1}5s_{1/2}$   & 2461.40 &     & 5.99[5]  & & 0.0005  & 0.004 \\
           & $\xrightarrow{M1}5p_{1/2}$   & 3651.94 &    & 4.43[-5] & & $\sim 0$ &  $\sim 0$   \\
           & $\xrightarrow{M1}5p_{3/2}$   & 3675.58 &    &  6.04[-3]& & $\sim 0$ & $\sim 0$ \\
           & $\xrightarrow{E2}5p_{3/2}$   & 3675.58 &   &  465.653 & & $\sim 0$  & $\sim 0$ \\
           & $\xrightarrow{E2}4f_{5/2}$   & 5371.75  &   & 43.43  & & $\sim 0$  & $\sim 0$ \\
           & $\xrightarrow{E1}5d_{3/2}$   & 13587.42&   & 2.69[7]  & & 0.372    & 0.170  \\
           & $\xrightarrow{E1}6s_{1/2}$   & 15883.73&   & 1.68[7]  & & 0.635  & 0.106  \\

$6p_{3/2}$ & $\xrightarrow{E1}3d_{3/2}$  & 642.78   &   &  6.41[6]   &  & 0.0004    & 0.040  \\
           & $\xrightarrow{E1}3d_{5/2}$  & 643.59   &    & 5.75[7]    & & 0.002  & 0.363 \\
           & $\xrightarrow{E1}4s_{1/2}$  & 769.02   &    & 2.68[7]    & & 0.005  & 0.169 \\
           & $\xrightarrow{M1}4p_{1/2}$  & 1069.85  &    &  0.003     &   & $\sim 0$ & $\sim 0$\\
           & $\xrightarrow{E2}4p_{1/2}$  &          &   & 837.868    &  & $\sim 0$ & $\sim 0$ \\
           & $\xrightarrow{M1}4p_{3/2}$  &  1075.3  & & 0.010  &          & $\sim 0$         & $\sim 0$ \\
           & $\xrightarrow{E2}4p_{3/2}$  &     &   & 1.32[3]    &  & $\sim 0$ & $\sim 0$ \\
           & $\xrightarrow{E1}4d_{3/2}$  & 2308.53    &    & 2.22[6]    &  & 0.002 & 0.014  \\
           & $\xrightarrow{E1}4d_{5/2}$  & 2310.95    &    & 1.99[7]    &  & 0.011 &  0.126 \\
           & $\xrightarrow{E1}5s_{1/2}$  & 2456.24    &    &  3.55[5]   &  & 0.0006 & 0.002  \\
           & $\xrightarrow{M1}5p_{1/2}$  & 3640.59    &   &  4.59[-3]  & & $\sim 0$ & $\sim 0$ \\
           & $\xrightarrow{E2}5p_{1/2}$  &    &  & 231.479   &  & $\sim 0$ & $\sim 0$ \\
           & $\xrightarrow{M1}5p_{3/2}$  & 3664.07  &   & 1.23[-3]   &  & $\sim 0$   & $\sim 0$   \\
           & $\xrightarrow{E2}5p_{3/2}$  &      &    & 2.31[2]    &  & $\sim 0$         & $\sim 0$  \\
           & $\xrightarrow{M1}4f_{5/2}$  & 5347.21    &  & 4.41[-11]     &  & $\sim 0$     & $\sim 0$   \\ 
           & $\xrightarrow{E2}4f_{5/2}$  &    &    & 6.21[1]     & & $\sim 0$  & $\sim 0$  \\
           & $\xrightarrow{E2}4f_{7/2}$  & 5347.28    &    &3.72[1]     & & $\sim 0$  & $\sim 0$   \\
           & $\xrightarrow{E1}5d_{5/2}$  & 13467.90   &    & 2.48[7]    && 0.451     & 0.157  \\
           & $\xrightarrow{E1}5d_{3/2}$  & 13431.53   &    &  2.76[6]   &   & 0.075  & 0.017 \\
           & $\xrightarrow{E1}6s_{1/2}$  & 15671.11   &   & 1.75[7]    &   & 1.289  & 0.110 \\
           & $\xrightarrow{M1}6p_{1/2}$  & 1170686.0  &   & 5.60[-6]   & & $\sim 0$ & $\sim 0$ \\
           & $\xrightarrow{E2}6p_{1/2}$  &       &   & 2.13[-9]   & & $\sim 0$  & $\sim 0$ \\
\end{longtable*}
References: $^a$ \cite{zeippen}; $^b$ \cite{ali}; $^c$ \cite{sahoo}; $^d$ \cite{wiese}.

In comparison to the above transition properties, it is observed that 
insufficient efforts are being made to accomplish any reliable results for
the lifetimes of different states in Sc III. Andersen et al \cite{andersen}
have measured lifetimes of the $4p$ states. In an outdated work, Buchta et 
al had carried out investigation of the lifetimes of a number of states
in the considered ion using a beam-foil technique measurement with 
reasonable accuracies \cite{buchta}. Some of the data reported by Wiese and 
Fuhr \cite{wiese}, as was mentioned in the previous paragraph, were actually
quoted from these measurements. We have estimated the lifetimes of all the
states that we have taken into account for our study using the transition 
rates given above and listed them in Table \ref{tab4} along side the
results of Andersen et al and Buchta et al. We have also estimated
uncertainties from the neglected Breit interaction and correlation effects
(slightly larger values are taken as upper limits) and they are quoted
inside the parentheses. Our estimated lifetimes for the $4f$ states are
completely disaccord with the results reported in \cite{buchta}. The cause for
the large discrepancies between these results could be due to the anticipated
error in the measurement as mentioned by Buchta et al in their paper; instead
these measurements may correspond to the lifetimes of the cascade $5g$ states.
We have also referred to few other theoretical estimations of the lifetimes of
the $4p$ states in the same table which were, in fact, determined from the 
mean-field theory based calculations. From the agreement between the
measured lifetimes for the $4p$ states with the mean-field theory results
using the velocity gauge expression than the length gauge expression, Buchta
et al have justified the accuracy of the mean-field based calculation and their
results \cite{buchta}. However, this agreement may be accidental, because as we
have stated in the earlier paragraph that the transition amplitudes from the DF
method (mean-field theory calculation) are usually larger in magnitudes compared
to the results obtained from the RCC method. So the mean-field results are
expected to give smaller lifetime values and the same was observed in the mean-field theory calculation. It is well known fact that the length gauge expression
gives faster converged amplitude result than the velocity gauge expression in an
approximated many-body theory calculations. Further, Buchta et al have given
other examples like how they find a similar observation for the estimation of
the lifetimes of the $4p$ states in the calcium ion (Ca II). We have also 
studied the lifetimes of various states using the length gauge expression in Ca 
II using our CCSD(T) method \cite{bijaya2,sahoo1} and shown the importance of 
the correlation effects to obtain results agreeing the with their corresponding 
measured values. Therefore, we think that the estimated lifetimes of the
$4p$ states for Sc III in this work are more precise than the measurements 
by Buchta et al \cite{buchta}. It would be appropriate to carry out further
measurements of the lifetimes of different states in this ion using the
the modern advanced experimental techniques to probe their accuracies and
test the potential of the many-body methods to reproduce them.

\begin{table}[t]
\caption{Lifetimes ($\tau$) of low-lying states in Sc III.} \label{tab4}
\begin{center}
\begin{tabular}{lccc}
\hline \hline \\
State & This work & Others & Experiments \\
\hline \\
  & \multicolumn{3}{c}{\underline{Lifetimes in $s$}}\\
$3d \ ^2D_{5/2}$ & 12135(100) & 12130.86$^a$ & \\
$4s \ ^2S_{1/2}$ & 0.05(1) & 0.0519$^a$ \\
 & & \\
  & \multicolumn{3}{c}{\underline{Lifetimes in $ns$}}\\
$4p \ ^2P_{1/2}$ & 1.43(2) & 1.6$^b$ & 1.7(2)$^{b,d}$ \\
$4p \ ^2P_{3/2}$ & 1.40(3) & 1.27/1.66$^c$ & 1.7(2)$^{b,d}$ \\
$4d \ ^2D_{3/2}$ & 0.95(1) &  & 1.2(2)$^d$ \\
$4d \ ^2D_{5/2}$ & 0.96(3) &  & 1.2(2)$^d$ \\
$5s \ ^2S_{1/2}$ & 1.08(2) & & 1.4(2)$^d$ \\
$5p \ ^2P_{1/2}$ & 3.32(2) &  & 3.6(4)$^d$ \\
$5p \ ^2P_{3/2}$ & 3.31(3) & & 3.6(4)$^d$ \\
$4f \ ^2F_{5/2}$ & 0.61(1) & & 3.5(8)$^d$ \\
$4f \ ^2F_{7/2}$ & 0.63(2) & & 3.5(8)$^d$\\
$5d \ ^2D_{3/2}$ & 2.56(1) & & 2.4(3)$^d$ \\
$5d \ ^2D_{5/2}$ & 2.63(1) & & 2.4(3)$^d$ \\
$6s \ ^2S_{1/2}$ & 1.66(1) & \\
$6p \ ^2P_{1/2}$ & 6.32(9) & \\
$6p \ ^2P_{3/2}$ & 6.33(8) & \\
\hline \hline \\
\end{tabular}
\end{center}
References: $^a$ \cite{sahoo}; $^b$ \cite{andersen}; $^c$ \cite{weiss}; $^d$ \cite{buchta}.
\end{table}
The lifetime and oscillator strength of the $5s$ state and $4p - 5s$ 
transition in Sc III were reported as 1.4(2)$ns$ and 0.13(2) in \cite{buchta},
which we obtain as 1.08(2)$ns$ and 0.168, respectively. Our oscillator strength
for the above transition match well with the tabulated result as 0.15 in
\cite{wiese}. Nonetheless, our results for the $5d$ states agree substantially
with the results reported by Buchta et al \cite{buchta}. The measured
oscillator strength for the $3d \rightarrow 4f$ transition is reported
as 0.03 which differ completely from our result 0.14, but our result 
agrees with the results reported in \cite{wiese}. With all the above
observations, it is worth reiterating that the results reported in this work 
are more accurate and they can be used reliably in any other applications.

 Using the forbidden transition amplitudes, we find very large lifetimes for 
the $3d \ ^2D_{5/2}$ and $4s \ ^2S_{1/2}$ states. The lifetime of the
$3d \ ^2D_{5/2}$ state is found to be 12135$s$ which is very large, because
of the highly forbidden transition between the corresponding fine structure
states (EE is very small). The lifetime of the $4s \ ^2S_{1/2}$ state found
to be 0.05$s$ which is also large enough in an atomic scale for carrying
out any other precision studies related to this state. These results are also
in perfect agreement with our previous findings \cite{sahoo}. 

\section{Conclusion}
We have employed the relativistic coupled-cluster method to determine
both the allowed and forbidden transition amplitudes in the doubly ionized 
scandium. By combining these results with the experimental wavelengths,
we have estimated the transition rates, oscillator strengths, branching
ratios and lifetimes for the first sixteen states in this ion. We have
compared our results with the previously reported ones and find a
reasonably agreement between them. The reported lifetimes of various 
states in this work seem to be meticulous than the previously available
results. Our results can be instrumental for various astrophysical studies
embodying scandium element. Further, these results can also be directive
for the new experiments to affirm the accuracies of the reported properties.

\section*{Acknowledgment}
B.K.S. acknowledges NSFC for the research grant no. 11050110417 and C.L. thanks
NSFC for the research grant no. 11034009. Computations were carried out using
3TFLOP HPC cluster at Physical Research Laboratory, Ahmedabad. 
Physical Research Laboratory, Ahmedabad.

\end{document}